\documentclass[newstyle,twocolumn,proceedings]{rmaa} 
\usepackage{rmaacite}
\title{When Photoionization is not Enough }  
\author{Sueli M. Viegas \affil{Instituto Astron\^omico e Geof\'\i sico, USP,
  Brasil}  }
\fulladdresses{  
\item S. M. Viegas: Instituto de Astron\^omico e Geof\'\i sico,  
  USP, Av. Miguel Stefano 4200, 04302-904 S\~ao Paulo, SP,   
  Brasil (viegas@iagusp.usp.br).}  
\shortauthor{Viegas}  
\shorttitle{When Photoionization is not Enough}  
\abstract{%
  It is well-known that photoionization is the major mechanism  
powering nebulae emission-lines. For about 30 years, numerical   
simulations using photoionization codes have been 
employed to analyze HII  
regions, planetary nebulae and active galactic nuclei. Although  
these models can reproduce most of the observed features,  
some remain unexplained. Two possible causes for  the   
discrepancies between observations and model predictions   
 are discussed in this paper: (a) the models are not  
realistic enough; (b) an additional energy source is needed.  
The discussion here is mainly based on the 
results of photoionization models  
 for the NW knot of I Zw 18.}   
\listofauthors{S.~M.~Viegas }  
\indexauthor{Viegas, S.~M.}  
\begin{document}  
\maketitle  
\section{Introduction}  
\label{sec:intro}  
The emission-line spectra of planetary nebulae, HII regions and 
active galactic nuclei have been largely used to derive
the physical conditions and elemental abundances in these
objects. An empirical method proposed by Peimbert \& Costero (1969) 
is usually used, providing average values of  the electron temperature and  
density of the emitting gas, as well as of the ionic fractions and   
elemental abundances, from the observed emission-line intensities. 
Since the main mechanism powering the emitting regions is photoionization,
numerical simulations based on photoionization codes have also
been used to study those objects. 
  
Since the 1960's, photoionization codes have been developed and   
 became a powerful tool to analyze the emission-line  
spectra of those objects. Presently, several such codes are  
available  (see, for instance,   
Ferland et al. 1995), one of the most popular being CLOUDY. 
Until recently,  
only these one-dimensional(1-D) codes, assuming    
spherical or plane-parallel symmetry, were available. In the  
last few years, some effort has been devoted to develop 3-D  
simulations (Gruenwald, Viegas \& Brogui\`ere 1997, Och, Lucy \& Rosa 1998).    
  
The 1-D photoionization codes have been used either to derive   
the general properties of a given class of emission-line objects or to  
obtain a detailed analysis of selected objects. In the first case,  
a grid of models is used to create diagnostic diagrams where  
theoretical and observational data are compared. On the other hand,  
detailed analysis of a given object, with observations available in   
a large wavelength range, provide a test for the physical processes  
occurring in the gas. However, even for HII regions 
and planetary nebulae,  which are surely powered by photoionization, 
the usual 1-D models fail  
to reproduce all their properties. The main problems are:   
  
(a) the observed  
radio brightness temperature is lower than predicted by the models. To  
 overcome  
this problem, it is usual to include a filling factor in the calculations as  
proposed by Osterbrock \& Flather (1959).  
   
(b) the  
discrepancy between the temperatures derived from the [O III] lines and from  
the Balmer lines is not explained by the models (see, for instance, Liu \&  
Danziger 1993). The discrepancy may be solved if the codes include  
either condensations (Viegas \& Clegg 1994) or    
additional heating to produce temperature fluctuations  
(Mathis, Torres-Peimbert \& Peimbert 1998 and references therein).   
  
(c) the discrepancy between the chemical abundances derived from  
forbidden and permitted emission-lines (Kaler 1986, Peimbert,  
Storey \& Torres-Peimbert 1993),  
 which could be explained by the presence of condensations    
(Viegas \& Clegg 1994, Liu et al. 2000).  
  
Regarding active galactic nuclei, self-consistent  
models reproducing both the  
emission-line and continuum spectra of selected objects indicate  
that shocks are present in the narrow-line region (Contini, Prieto   
\& Viegas 1998a,b, Contini \& Viegas 1999).  
  
In the following, the main issues associated with photoionization  
models are reviewed. In particular, models for the  
 well-known HII galaxy, I Zw 18, are discussed, and possible explanations  
for the discrepancy between model results and observational data  
are suggested.   
  
\section{Star forming galaxies}   
\label{sec:star}  
Due to its low metalicity, I Zw 18 is one of the extragalactic HII regions  
used to derive the primordial He abundance.   
Recently, Stasinska \& Schaerer (1999, hereinafter SS99)   
presented 1-D models for  
 the NW knot of  I Zw 18. They concluded that in addition to photoionization,  
another heating mechanism is necessary to explain all 
 observed features. On the other hand, in starburst 
galaxies, shocks must be contributing to the 
observed emission-lines depending on the evolutionary phase 
of the stellar cluster (Viegas, Contini \& Contini 2000). 
Could this be the case  for I Zw 18?  In order to answer this 
question,  photoionization models for I Zw 18 are reexamined.  
  
\subsection{I Zw 18: the SS99 model}  
  
A 1-D spherically symmetric photoionization model 
for the NW knot of I Zw 18  
was proposed by SS99, assuming the ionizing radiation   
spectrum of a stellar cluster, a  
density of 100 cm$^{-3}$ (as indicated by the [S II] line ratio) 
and chemical   abundances derived by Izotov \& Thuan (1998).
In defining the  
best fit, the criteria adopted were that the relevant emission-line ratios   
must be reproduced, as well as  
the size of the ionized region and its luminosity.   
  
Adjusting the total initial mass of the stellar 
cluster in order to reproduce the luminosity,  the authors  
show that a homogeneous model may explain some of the line ratios, but   
the calculated size of the ionized region is too small unless  
a filling factor of the order of 10$^{-2}$ is adopted. 
Notice that the HST image of I Zw 18 (see figure 1 of SS99)  
 shows an inhomogeneous gas distribution which could  
justify models with a filling factor less than unity.
The homogeneous model with $\epsilon$ = 10$^{-2}$ combined  
with condensations provided the best-fit. All the criteria
were fullfilled except for  
the [O III] $\lambda$4363/5007 line ratio, which was too low
compared to observations. Since this  
line ratio is a known temperature indicator, the  
authors concluded that an additional heating mechanism must be present  
in the NW knot of I Zw 18.    
  
\subsection{Shocks in Star Forming Regions}  
  
The detection of the infrared line [O IV]25.9 $\mu$m, in addition to  
[Ne II]12.8 $\mu$m and [Ne III]15.6$\mu$  in a number of  
well-known starburst galaxies has been reported by Lutz et al. (1998).  
Using simple photoionization  
models, these authors showed that   
photoionization by a stellar cluster was not enough to explain  
the presence of the highly ionized species. They suggested other  
possible mechanisms that could contribute to the emission-lines. They   
concluded that  
shocks related to  stellar activity must prevail in these objects.  
However, it was shown  that the presence of hot Wolf-Rayet stars  
could explain the observed [O IV] emission for two galaxies  
of the Lutz sample: NGC 5253 and II Zw 40 (Schaerer \& Stasinska 1999).  
But even considering WR stars, 
photoionization models are unable   
to explai the observations of the other sample galaxies, leading   
Schaerer \& Stasinska to conclude that an alternative 
mechanism is required.
   
To analyse the galaxies of the Lutz sample, 
Viegas, Contini \& Contini (1999) used
composite models, coupling the effect of photoionization 
by a stellar cluster and shocks.  
  
The numerical simulations were performed using the SUMA code
 (see, for instance, Viegas \& Contini 1997). This code has been  
used since the 1980's to study the narrow-line regions of active galactic nuclei. The   
ionization and heating equations are solved for a plane-parallel   
cloud, taking into account the  
coupled effect of ionizing radiation  and  shocks, and  
also accounting for the diffuse radiation produced by the hot gas.  
Due to this self-consistency, SUMA models are  more realistic than  
those obtained by the so-called ionizing  
shock models, which only account for the diffuse radiation
generated by the hot gas. However  it is well known that  
a radiation source is always present in AGN, planetary nebulae  
and HII regions, and although its effect on the 
physical conditions of the gas may
be less important than shocks, it is hardly negligible and cannot be 
turned off.
    
There are two types of composite models depending on the  
location of the cloud relative to the radiation source and  
the shock front. In one case, both ionizing  
mechanisms  affect the same edge of the cloud, whereas  
in the other case,  
they act on opposite edges. In the latter case,
 the effects of photoionization and shocks may  
overlap through the diffuse radiation, provided  
the cloud is not too wide. Regarding the simulations,  
in addition to the input parameters usually used 
in photoionization models  
(ionizing radiation spectrum, density distribution and chemical  
abundances), the composite models requires the shock velocity,  
the pre-shock density and the geometrical thickness of the cloud.  

\begin{figure}  
\includegraphics[width=\columnwidth]{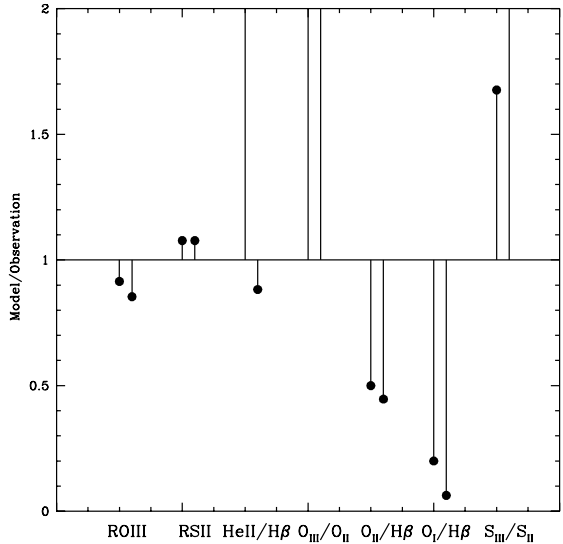}  
\caption{The size of the ionized gas (R$_f$/10$^{19}$)  
and the gas temperature (T/10$^4$) as a function of the filling factor  
adopted in the photoionization model. The dotted line  
indicates the observed value of the radius, whereas the dashed lines  
indicate the range associated with the gas temperature   
deduced from the [O III]  line ratios.  
Although observations are not available for the [N II] line ratio,  
the corresponding temperature obtained from the models is also  
plotted.}  
\end{figure}  
  
Assuming a black-body spectrum, and a range of shock velocities and
pre-shock densities, 
single-cloud models have been built for the starburst galaxies of 
Lutz sample. The results are
compared to the observed infrared line ratios using a diagnostic 
diagram [O IV]/([Ne II] + 0.44[Ne III]) versus [Ne III]/[Ne II],
 as suggested by Lutz et al. (1998).  
In the diagram, the data points are distributed along a diagonal,  
going from the bottom-left corner to the  
top-right corner. The points corresponding to the two galaxies 
NGC 5253 and II Zw 40,   
studied by Schaerer \& Stasinska (1999), are located 
at the top/right zone.   
   
Assuming that all the  
objects are photoionized by a similar radiation   
spectrum,  our single-cloud results show that the distribution  
on the diagram can be reproduced if the shock velocity and the  
pre-shock density increase from the top to the bottom part of the diagonal.   
When the intensity of ionizing radiation is kept constant, the  
higher the shock velocity the stronger is the shock effect  
on the physical conditions in the cloud,  
relative to photoionization. Since NGC 5253 and II Zw 40 data points  
are located in the low velocity --- low density zone, photoionization  
 is expected to be the dominant mechanism for these two objects,  
while shocks are required for explaining the data for the other objects, in agreement with Lutz et al. (1998).   

\begin{figure}  
\includegraphics[width=\columnwidth]{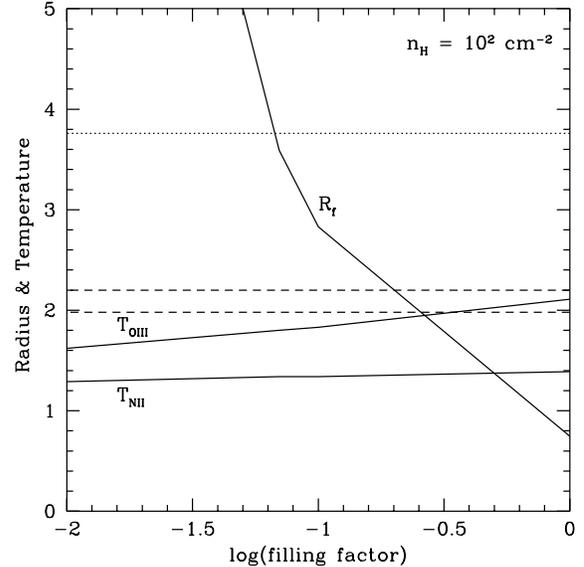}  
\caption{Comparing observational and theoretical intensity ratios  
of the relevant emission-lines: ROIII = [O III]4363/5007,   
RSII = [S II]6717/6731, HeII4686/H$\beta$,   
[O III]5007+4959/[O II]3727, [O II]3727/H$\beta$,   
[O I]6300+6363/H$\beta$ and [S III]6312/[S II]6727+6731.  
The results correspond to a uniform density cloud (30 cm$^{-3}$),  
photoionized by a stellar cluster. For each line ratio,   
the symbols on the left correspond to a 3.3 Myr stellar cluster,  
while those on the right to a 5.4 Myr stellar cluster.}   
\end{figure}

From our results, the observed trend can be interpreted as a sequence  
between two extreme cases: starburst regions, where high velocity shocks  
are present and give a relevant contribution to gas ionization,  
and radiation-dominated objects (HII galaxies),   
where photoionization is the dominant mechanism. As showed  
by Viegas et al. (1999), the distinction between  
these two kinds of galaxies can be understood in terms of temporal  
evolution. In HII galaxies, like NGC 5253 and II Zw 40, the star  
forming region is young enough to contain a high fraction of   
massive O and WR stars which produce a large amount of ionizing photons.  
On the other hand, in starburst galaxies such
 as M 82 and NGC 253, the  
star formation event is much older and the massive stars have already  
evolved to a SN phase. As the stellar cluster evolves, 
the ratio between the mechanical  
energy from the SNe and the ionizing Lyman continuum rapidly increases
(Leitherer \& Heckman 1995). Thus, radiation dominated HII 
galaxies must be associated with young metal poor objects, whereas  
starburst galaxies with massive evolved objects.

Since I Zw 18 has the characteristics of a HII galaxy, we expect   
photoionization to be the major ionizing mechanism. Thus,   
in the following we review the photoionization model presented  
by SS99, which indicated the need for an  
additional heating mechanism.  
  
\section{Reviewing the NW knot of I Zw 18}  
\label{sec:izw18}  
    
\begin{figure}  
\includegraphics[width=\columnwidth]{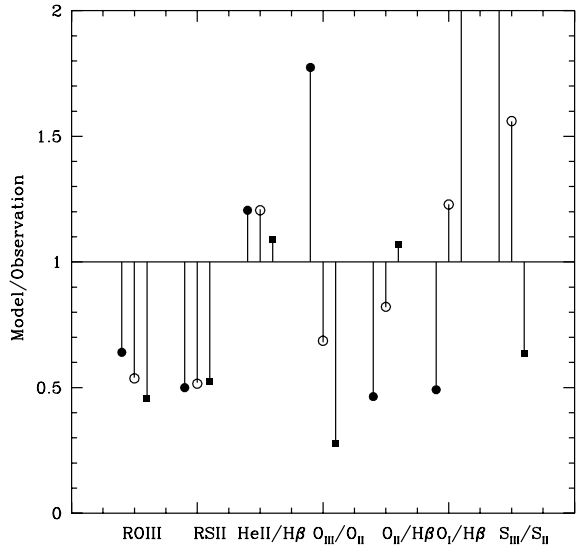}  
\caption{As shown in Figure 2 for a high-density 
filament (10$^4$ cm$^{-3}$)  
photoionized by a 5.4 Myr stellar cluster.  For each line ratio,    
three results are shown corresponding to different locations of the  
filament in the HII region: at the nebula inner edge (solid dots),   
and at  distances from the   
stellar cluster where 25\% (solid squares) and 75\% (circles)  of   
H$\beta$ has already been emitted.}  
\end{figure}  
  
The suggestion of  an additional ionizing mechanism   
in the NW knot of I Zw 18 came from the fact that 
the photoionization models  
proposed by Stasinska \& Schaerer (1999)  
were unable to reproduce the [O III] line ratio. This line ratio is   
a temperature indicator, and their models gave an 
average gas temperature  
too low compared to that derived from the observations.   
  
To analyze the problem, the model assumptions are   
reviewed and a new model is proposed using  
 the 1-D  photoionization code Aangaba (Gruenwald  
\& Viegas 1992). In the following, the initial models  
are built assuming the same input parameters  
adopted by SS99, i.e., n$_H$ = 100 cm$^{-3}$,  
 a stellar cluster ionizing continuum and the
Izotov \& Thuan (1998) chemical abundances.  
  
The results  shown in Figure 1   
correspond to a spherically symmetric nebula  
photoionized by a continuum due to a 3.3 Myr old stellar cluster  
(Cid-Fernandes et al. 1992). The total mass is  
chosen in order to reproduce the observed luminosity. It is easily  
verified that, unless the filling factor   
is of order   
10$^{-2}$, the size of the ionized region is too small.  
  
On the other hand, as $\epsilon$ approaches unity,  
the gas temperature defined by the [O III] ratio becomes higher.  
Notice that photoionization  
models can match the observed gas temperature T$_{OIII}$  
as long as the adopted filling factor is higher than 0.3.  
In this case, however, the calculated outer radius is far below the  
observed value.   
  
The net effect of 1-D models with  $\epsilon$    
lower than unity is to  provide a slower increase of the  optical   
depth  with the distance  from the central source (see, for instance,  
Osterbrock 1989). This  leads to an 
increase of the size of the low-ionization zone  
with respect to the high-ionization zone, which results in   
lower average temperatures derived from more ionized  
species, as in the case of the [O III] lines.

A possible way to increase the ionized zone 
without decreasing the average  
gas temperature is to assume  $\epsilon$ = 1 and a  
gas density lower than 100 cm$^{-3}$. This value was adopted  
by SS99 in order to reproduce the observed [S II] line ratio  
from Izotov \& Thuan (1998). However, it is well-known that  
the [S II] line ratio is practically constant for densities  
lower than a few hundred. Therefore, models with a density  
of the order of 30 cm$^{-3}$ and  $\epsilon$ = 1 can reproduce  
both the size of the ionized region and the  T$_{OIII}$ temperature.  
The emission-line ratios relevant to constrain this model are  
shown in Figure 2, where  observed values are compared to  
the values coming from a uniform gas distribution with  
density equal to 30 cm$^{-3}$,  $\epsilon$ = 1, and assuming  
ionization from stellar clusters of two different ages: 
3.3 Myr and 5.4 Myr.  
The latter gives a better fit for HeII/H$\beta$.  
On the other hand,  [OIII]/[OII] and [SIII]/[SII] are too high,  
while [O II]/H$\beta$ and [O I]/H$\beta$ are too low.  
  
The same kind of discrepancy between observational and theoretical  
line ratios was reported by SS99 concerning the homogenous model.  
These authors suggested the presence of condensations to avoid  
the fitting problems. In Figure 3, results are shown for  
filaments with density equal to 10$^4$ cm$^{-3}$, located   
at different distances from a stellar cluster. 
Notice that the results can be very different. A real model  
for the NW knot of I Zw 18 should account for the contributions  
of the diluted gas as well as of the different filaments. 
It is clear that a model can be obtained with the same degree of
accuracy as obtained by SS99 but with the additional
advantage of also fitting the [O III] line ratio.

However, more than finding the best model for I Zw 18,  
the goal here is to illustrate how assumptions used to build a model
that satisfies practically all the observational constraints,  
could still lead to a wrong conclusion. Indeed, models with a filling  
factor lower than unity and higher density (100 cm${-3}$)
 indicated that an additional  
energy source was required, while models with $\epsilon$ = 1
and lower density (30 cm${-3}$) offer a fitting of the 
line ratios, H$\beta$ luminosity and size of the ionized region without that requirement. Thus, 
pure photoionization can explain I Zw 18 observations, as expected  
from the analysis of  
the infrared lines of star forming objects (Viegas et al. 1999).

\section{Concluding Remarks}  
\label{sec:concluding}  
The improvement of spatial resolution,  
revealing  more and more details of the emitting regions of  
different classes of objects,  
showing the presence of clumps and voids, as well as the   
improvement in signal/noise,  providing more   
accurate line intensities, require a more realistic approach.   
  
It seems clear to me  
that if 1-D models are used to analyze the emission-line  
spectra of  clumpy objects, the use of a filling factor  
less than unity should be avoided. As shown in the case of  
the NW knot of I Zw 18, in this kind of model the  
conclusions may be misleading because the effect of $\epsilon$  
$<$ 1 is to artificially increase the low-ionization zone  
of the cloud. As a consequence, optically thick models will  
artificially decrease the average temperature of the 
high-ionization zone (see Fig. 1) and artificially increase the  
intensity of the low ionization lines. Thus, the chance of  
reaching a wrong conclusion  about the gas temperature  
or the ionic fraction low-ionization species is very high.  
In the first case, the presence of additional heating source   
required by the model may not be necessary, as 
illustrated by the discussion of I Zw 18.  
In the second case, mimicking a higher ionic fraction for  
the low-ionization species may reflect both on the elemental  
abundances deduced from the model, as well as on the   
choice of the other input parameters. A good example 
 is the model used by Alexander et al. (1999) to  
study the ionizing continuum of NGC 4151. Their best-fit model  
for the infrared lines   
corresponds to a filling factor $\epsilon$ $<$ 1. Based on this model  
they discard the existence of a big blue bump just beyond  
the Lyman limit because ``{\it such a bump clearly  
overproduces the low-ionization lines and underproduces  
the high-ionization lines}''. Notice however that this could  
also be a consequence of the filling factor adopted by the authors.  
  
A better and safer way to study clumpy regions with 1-D models was  
proposed more than a decade ago by 
Viegas-Aldrovandi \& Gruenwald (1988)  
when studying the narrow-line regions of AGN. Several models with  
different physical conditions are combined to produce the  
observed emission-line spectra. This method has also been  
used by Contini et al. (1998a,b, 1999) to reproduce the emission-line  
and continuum spectra of Circinus, NGC 5252 and NGC 4051  
 self-consistently.  
  
Certainly a better approach to emission-line objects  
 is to use a 3-D photoionization model,  
specially if imaging and spectroscopic data with high spatial  
resolution are available (Morisset et al. 2000).   
A good example is the planetary nebula NGC 3132. Its imaging  
seems to indicate that it has ellipsoidal geometry.  
However, using a 3D code,  Monteiro et al. (2000, 2001) showed that
a model with an  ellipsoidal geometry could not reproduce all  
the observational data. In fact, only a hourglass 
(diabolo) shape could provide a self-consistent fit   
to the image, the emission-line spectrum, the electron  
density distribution and the velocity field. There is no doubt that  
3-D models will become the  tools of choice
analyzing emission-line objects.  
  
Based on all the discussion above, let me end with  a warning:   
{\it When dealing with models, beware of possible illusions...}

\acknowledgements I am  very grateful to my friends   
Ruth Gruenwald, Marcella Contini and M. Almudena   
Prieto for a long and fruitful collaboration,  
which permitted me to present this paper.  I am also indebt
to Gary Steigman for a careful proofreading.
This work has been partially supported by PRONEX/Finep(41.96.0908.00)  
and CNPq (304077/77.1)

\end{document}